\documentstyle[11pt,aaspp4,epsfig]{article}

\begin{document}

\title{Phase Lags of QPOs in Microquasar GRS 1915+105}
\author{Wei Cui}
\affil{Center for Space Research, Massachusetts Institute of Technology, 
Cambridge, MA 02139; cui@space.mit.edu}

\begin{abstract}
I report the discovery of hard X-ray phase lags associated with the 
famous 67 Hz quasi-periodic oscillation (QPO) in microquasar GRS 
1915+105. The QPO is seen on multiple occasions. For this
investigation, I have chosen one particular observation when the 
oscillation is the strongest. The measured hard lags show strong 
energy dependence. With respect to the 2-2.5 keV band, the phase lag 
increases from insignificant detection at 5.2-7.0 keV to as much as 
2.3 radians (which corresponds a time lag of $\sim$5.6 ms) above 
$\sim$13 keV. Also detected in the same observation are one narrowly 
peaked, strong QPO at 67 mHz, along with its first three harmonics, 
one weak QPO at $\sim$400 mHz and one broad QPO at $\sim$840 mHz. 
Similar cross spectral analyses have been performed on these QPOs 
(and the harmonics), in order to measure any phase lags associated 
with the features. Phase lags are detected in all, with similar 
energy dependence. For the 67 mHz QPO, the results are quite 
intriguing: only the first and third harmonic components show hard 
lags, while the fundamental and the second harmonic components 
actually display smaller {\it soft} lags. Coupled with the energy 
dependence of the QPO amplitude, the results seem to indicate a 
complicated change in the signal profile with photon energy. I will 
discuss the implication of the phase lags on possible origins of the 
QPOs.
\end{abstract}


\section{Introduction}
Black hole candidates (BHCs) are known to vary strongly in X-ray
(van der Klis 1995; Cui 1999). The variability can sometimes show
a characteristic periodicity, in the form of a quasi-periodic 
oscillation (QPO). For BHCs, QPOs were initially observed only in 
a few sources at very low frequencies ($<$ 1 Hz), with the exception 
of the rare ``very-high-state'' (VHS) QPOs at a few Hz observed of 
GX 339-4 and GS 1124-68 (van der Klis 1995 and references therein). 
Since the launch of Rossi X-ray Timing Explorer (RXTE; Bradt et al. 
1993), new QPOs have been discovered at an accelerated pace, thanks 
to the advanced instrumentation of RXTE. Not only are the QPOs now 
seen in more BHCs, they are also detected at increasingly higher 
frequencies (Cui 1999 and references therein). The phenomenon now 
spans a broad spectrum from a few mHz to several hundred Hz. Despite 
the observational advances, the origin of QPOs remains uncertain.
Progress has been made empirically by correlating the observed 
properties of the QPOs, such as centroid frequency and amplitude, to 
physical quantities, such as photon energy and X-ray flux (or mass 
accretion rate). It has been shown that the correlations can be quite 
different for different QPOs, perhaps indicating that for BHCs the 
QPOs form a heterogeneous class of phenomena (Cui 1999; Cui et al. 
1999a).

GRS 1915+105 is one of only three known BHCs that occasionally produce
relativistic radio jets with superluminal motion (Mirabel \& Rodriguez
1999 and references therein). These sources are often referred to as
microquasars. First discovered in X-ray (Castro-Tirado et al. 1992), 
GRS 1915+105 has been studied 
extensively at this wavelength. Recent RXTE observations revealed a 
great variety of QPOs associated with the source (Morgan et al. 1997), 
in addition to its complicated overall temporal (as well as spectral)
behaviors (e.g., Greiner et al. 1996). The most famous of all is the 
QPO at 67 Hz. At the time, it was the highest-frequency QPO ever
detected in BHCs. More interestingly, the centroid frequency of
the QPOs hardly varies with X-ray flux (Morgan et al. 1997), unlike
a great majority of other QPOs. Suggestions have subsequently been 
made to associate the feature to the dynamical processes in the 
immediate vicinity of the central black hole, where general
relativistic effects may be strongly manifested (Morgan et al. 1997; 
Nowak et al. 1997; Cui et al. 1998). As a result, a lot of excitement 
has been generated by the prospect of using such signals to test the 
general theory of relativity in the strong-field limit (Cui et al. 
1999b and references therein). Before this ultimate goal can be
reached, however, it is clearly important to best characterize and 
understand this particular QPO observationally. In this Letter, I 
report the discovery of an important property of the feature: the 
oscillation lags more behind at higher photon energies. Some of the 
results have already been presented elsewhere in preliminary 
form (Cui 1997; Cui 1999).

\section{Observations}
The 67 Hz QPO has been detected in various different states of GRS
1915+105 (Morgan et al. 1997). For this investigation, I have selected
one RXTE observation when the oscillation appears the strongest (based 
on Morgan et al. 1997). The observation was made on May 5, 1996, when 
the source was in a bright state, with a total exposure time about 10 
ks. Multiple high-resolution timing modes were adopted for
the observation. Considering the trade-off between statistics and
energy resolution, I have decided to rebin the $16\mu s$ {\it Event} 
data to $2\mbox{ }ms$ and to combine the sixteen energy bands (above 
$\sim$13 keV) into one. I have then merged the {\it Event} mode data 
with the $2\mbox{ }ms$ {\it Binned} mode data (which covers four
energy bands below $\sim$13 keV). Now, a total of 5 energy bands are 
available for carrying out subsequent analyses. The bands are 
approximately defined as 2--5.2 keV, 5.2--7.0 keV, 7.0--9.6 keV, 
9.6--13.2 keV, and 13.2--60 keV.

\section{Data Analysis and Results}
A collection of power-density spectra (PDS) of GRS 1915+105 can be
found in Morgan et al. (1997), for the initial 31 RXTE observations of 
the source, but only in one energy band (2--20 keV). Of great interest 
here is the energy dependence of the temporal 
properties of the source, so I have constructed the PDS (with the 
deadtime-corrected Poisson noise power subtracted) in the five energy 
bands defined. Fig.~1 shows the results. The presence of QPOs is 
apparent. Most prominent is the 
one centered at about 67 mHz, along with its first three harmonics. As 
noted by Morgan et al. (1997), the amplitude of the fundamental
component seems to follow a general decreasing trend toward high 
energies, while that of the harmonics shows just the opposite. The 67 
Hz QPO is clearly visible in the PDS, especially at high energies. 
Also detected are a weak QPO at about 400 mHz (which is missing from 
Morgan et al. 1997) and a broad QPO at about 840 mHz.

To be more quantitative, I have fitted the PDS with an empirical model
consisting of Lorentzian functions for the QPOs and a double-broken 
power law for the underlying continuum. I have limited the frequency 
range to roughly 0.001--10 Hz during the fitting to focus on the 
low-frequency QPOs, since the 67 Hz QPO has already been quantified 
(Morgan et al. 1997). Table~1 summarizes the best-fit centroid 
frequency and width for each QPO at low frequencies, derived from the 
7.0--9.6 keV band (which is chosen as a compromise between the signal 
strength and the data quality). Fig.~2 shows the fraction 
rms amplitude of each QPO at different energies. The 400 mHz QPO and
the 840 mHz QPO, as well as the 67 Hz QPO (Morgan et al. 1997),
strengthen toward high energies (with a hint of saturation), similar 
to most QPOs of BHCs (Cui 1999). The behavior of the 67 mHz QPO is, 
however, more complicated and very intriguing: while the 
harmonics of the QPO follow the usual trend, the fundamental component 
becomes stronger first and then weakens significantly at higher 
energies, with the amplitude peaking at 8--9 keV.

To derive phase lags, I have chosen the 2--5.2 keV band as a reference
band, and computed a cross-power spectrum (CPS) between it and each
of the higher energy bands. Note that the final CPS represents an 
ensemble average over the results from multiple 512-second segments of 
the time series (similarly for the PDS shown). The phase of the CPS 
represents a phase shift of the light curve in a selected energy
band with respect to that of the reference band. Here, I follow the 
convention that a positive phase indicates hard X-rays lagging behind 
soft X-rays, i.e., a hard lag. The uncertainty of the phase is
estimated from the standard deviation of the real and imaginary parts 
of the CPS. The magnitude of the QPO lags is derived from fitting the 
profile of the 
lags with Lorentzian functions at the QPO frequencies which are fixed 
during the fitting. Note that for the 67 Hz QPO I have also fixed the 
width of the profile to that of the QPO (see Table~2 in Morgan et al. 
1997), due to the lack of statistics. The measured lags are plotted 
in figures~3 and~4 for the 67 Hz QPO and 
other low-frequency QPOs, respectively. The errors are derived by 
varying the parameters until $\Delta \chi^2 = 1$ (i.e., corresponding 
roughly to $1\sigma$ confidence intervals; Lampton et al. 1976). 

Most QPOs show significant hard lags. Surprisingly, however, the 
odd harmonics of the 67 mHz QPO display {\it soft} lags. It is clear 
that the QPO lags depend strongly on photon energy --- the higher the
energy the larger the lag, with the exception of the 840 mHz QPO where 
the measured hard lag increases first and then drops above 13 keV. 
For the 67 Hz QPO, the phase lag reaches as 
high as 2.3 radians, which is equivalent to a time lag of about 5.6
ms. The phase lags are smaller for low-frequency QPOs, but the 
corresponding time lags are quite large. For instance, the first 
harmonic of the 67 mHz QPO shows a time lag greater than 1 second for 
the highest energy band. 

\section{Discussion}
It is known that hard lags are associated with the X-ray emission from 
BHCs (van der Klis 1995; Cui 1999). Although the studies of hard lags 
are mostly based on broad-band variability, the large lags associated 
with the VHS QPOs of GS 1124-68 have been noted (van der Klis 1995). 

Often, the hard lags are attributed to thermal inverse-Comptonization 
processes (e.g., Miyatomo et al. 1988; Hua \& Titarchuk 1996; Kazanas 
et al. 1997; B\"{o}ttcher \& Liang 1998; Hua et al. 1999), which are 
generally thought to be responsible for
producing the hard power-law tail of X-ray spectra of BHCs (Tanaka \& 
Lewin 1995). In these models, the lags are expected to be larger for 
photons with higher energies, since a greater number of scatterings 
are required for seed photons to gain enough energy. More 
quantitatively, the hard lags, which indicate the diffusion timescales 
through the Comptonizing region, should scale logarithmically with 
photon energy (e.g., Payne 1980; Hua \& Titarchuk 1996); this roughly 
agrees with the observations (Cui et al. 1997; Crary et al. 1998; 
Nowak et al. 1999; also see figures~3 and~4). 
However, the measured time lags can often be quite large, e.g., a few 
tenths of a second, at low frequencies, which 
would require a large hot electron corona (roughly one light second
across; Kazanas et al. 1997; B\"{o}ttcher \& Liang 1998; Hua et
al. 1999). A even larger corona would be needed to account for the
hard lags observed of the first harmonic of the 67 mHz QPO in GRS
1915+105. It is still much debated whether the required corona can be 
maintained physically (Nowak et al. 1999; B\"{o}ttcher \& Liang 1998; 
Poutanen \& Fabian 1999). Also, the observed soft lags of the 67 mHz 
QPO (and its second harmonic) are entirely incompatible with the
Compton models. On the other hand, the smaller time lags associated
with the QPOs at higher frequencies (e.g., the 67 Hz QPO) can still be 
accommodated by the models. Like the QPOs themselves, the phase lags 
may very well be of multiple origins.

Another class of models link the time lags to the propagation or drift
time scales of waves or blobs of matter through an increasingly hotter 
region, toward the central black hole, where hard X-rays are emitted
(Miyamoto et al. 1988; B\"{o}ttcher \& Liang 1998). In this scenario,
as the disturbance (such as waves, blobs, and so on) propagates, its 
X-ray spectrum hardens, producing the observed hard lags. The models
can also produce the logarithmic energy dependence of the hard lags 
(B\"{o}ttcher \& Liang 1998). The origin of clumps of matter may lie 
in the Lightman-Eardley instability which sets in wherever radiation 
pressure dominates over gas pressure in the accretion disk (Lightman 
\& Eardley 1974). It has recently been proposed that such a condition 
is perhaps generally met for the inner region of accretion disks in 
BHCs and thus the presence of blobs may not be surprising (Krolik
1998). However, it is not clear how to associate 
QPOs with the dynamics of the blobs. While the Keplerian motion of the
blobs could manifest itself in a QPO observationally, any radial drift 
of the blobs would cause the QPO frequency to increase with energy, 
which is not observed. 

On the other hand, it is perhaps easier to associate QPOs with traveling 
waves. For instance, Kato (1989) suggested that the propagation of 
corrugation-mode oscillations might explain the hard lags observed of
BHCs. Note also that inward-propagating perturbations have recently
been invoked to explain the X-ray variability of BHCs (Manmoto et al. 
1996), in the context of advection-dominated accretion flows (ADAFs; 
e.g., Narayan \& Yi 1994). In general, recent works seem to converge 
toward an ADAF-like geometry for the accretion flows around black
holes, i.e., an inner quasi-spherical, optically thin region and 
an outer thin, optically thick disk (e.g., Narayan \& Yi 1994; Chen et 
al. 1995; Dove et al. 1997; Luo \& Liang 1998). Applied to GRS
1915+105, the wave-propagation models might be able to account for the 
hard lags observed of the QPOs,
although the required speed of propagation would be quite small (much 
smaller than that of free fall) in some cases. Moreover, the peculiar
behavior of the 67 mHz QPO --- the even harmonics showing soft 
lags while the odd harmonics hard lags --- can perhaps be explained by
invoking a significant change in the wave form as the wave propagates.
To better illustrate this point, I have simulated an oscillation of
fundamental frequency 67 mHz that also includes the first three of its
harmonics. A sine function is used to describe each harmonic
component, which uses the 
measured fractional rms amplitude and phase of each component of the 
67 mHz QPO in GRS 1915+105. The overall profiles of the simulated 
oscillation are constructed by summing up the four components for the 
2--2.5 keV band (the reference band in which the phases are
initialized to zero) and the 13.2--60 keV band, respectively. They are 
shown in Fig.~5. The inferred evolution of the oscillation 
profile is quite drastic. The soft and hard lags are mixed together 
in the figure, contributing to the overall variation of the wave
form. But, because the hard lag is so dominating, it can be still be 
recognized by comparing the times of the minimum points between the two
profiles. However, the models 
cannot naturally explain why the fractional amplitude of the QPOs 
increases with energy if the QPOs are of disk origin (Cui 1999);  
neither can most Compton models, unless a certain spatial distribution
of the Compton y-parameter is assumed (Wagoner et al. 1999; Lehr et al. 
1999).

A third class of models associate the hard lags with dynamical
processes in Comptonizing regions themselves. Poutanen \& Fabian
(1999) proposed that the time lags may be identified with the spectral
hardening of magnetic flares as the magnetic loops inflate, detach and 
move away from the accretion disk. Now, the time lags are directly
related to the evolution timescales of the flares. This model differs
fundamentally from others in allowing Comptonizing regions to vary. 
It, therefore, provides an interesting possibility that the QPOs of
BHCs might be an observational manifestation of the oscillatory nature 
of these regions. The oscillation may occur, for example, in the 
temperature of hot electrons and/or in the Compton optical depth, both 
of which cause ``pivoting'' of the Comptonized spectrum. The spectral
pivoting might naturally explain the observed energy dependence of the 
QPO amplitude (e.g., Lee \& Miller 1998; see Kazanas \& Hua 1999 for 
another possibility).

To summarize, besides other observable properties of the QPOs, the phase
lags provide additional information that may be critical for our
understanding of the QPO origins. The energy dependence of the lags has 
already had serious implications on theoretical models ---
Comptonization seems always required. It might also shed light on the 
evolution of intrinsic wave forms, when combined with the energy 
dependence of the QPO amplitude, and thus on underlying physical 
processes and conditions that can cause the evolution. Moreover, the 
magnitude of the lags might provide a direct measure of such important 
physical properties of the system as the size of the Comptonizing 
region, or the propagation speed of disturbances in accretion flows, 
or the evolution timescales of magnetic flares originating in the 
accretion disk. Therefore, the QPO lags might ultimately prove 
essential for understanding the geometry and dynamics of mass
accretion processes in BHCs.  

\acknowledgments
I gratefully acknowledge useful discussions with many participants of
the first Microquasar Workshop (which was held in May, 1997 at the 
Goddard Space Flight Center, Greenbelt Maryland) where some of the
preliminary results were first presented. I thank the referee, 
Dr. Bob Wagoner, for his prompt report and many useful comments.
Financial support for this work is partially provided by NASA through 
an LTSA grant and several RXTE grants.

\clearpage

\begin{deluxetable}{ccc}
\tablecolumns{3}
\tablewidth{0pc}
\tablecaption{Low-frequency QPOs in GRS 1915+105}
\tablehead{
\colhead{Frequency (mHz)}&\colhead{FWHM (mHz)}&\colhead{Comments}}
\startdata
$66.8^{+0.1}_{-0.1}$ & $1.2^{+0.2}_{-0.2}$ & fundamental component \nl
$133.3^{+0.2}_{-0.2}$ & $2.4^{+0.4}_{-0.4}$ & first harmonic \nl
$201.2^{+0.3}_{-0.6}$ & $4.0^{+0.9}_{-1.0}$ & second harmonic \nl
$267.7^{+0.6}_{-0.7}$ & $7.4^{+1.1}_{-1.0}$ & third harmonic\nl
$403^{+2}_{-2}$ & $14^{+8}_{-6}$ & \nodata \nl
$841^{+5}_{-5}$ & $535^{+19}_{-18}$ & \nodata \nl
\enddata
\label{tb:qpo}
\end{deluxetable}

\clearpage
\begin{figure}
\psfig{figure=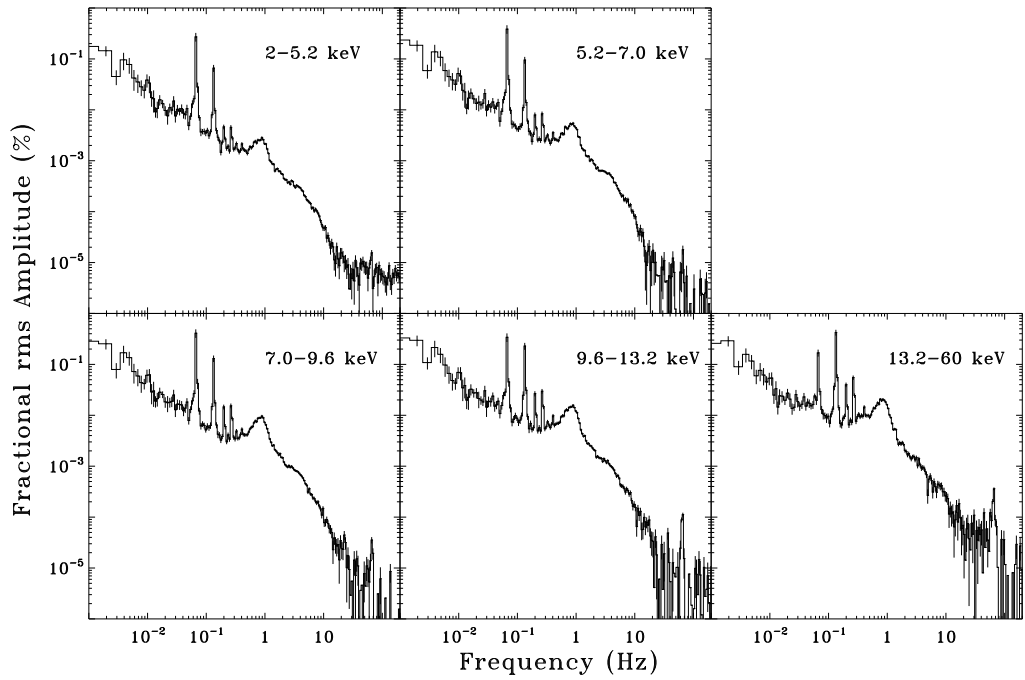,width=4in,angle=90}
\caption{Power density spectra of GRS 1915+105 in five energy
bands. Note that the dead-time corrected noise power due to counting 
statistics has been subtracted. }
\label{fg:pds}
\end{figure}

\begin{figure}
\psfig{figure=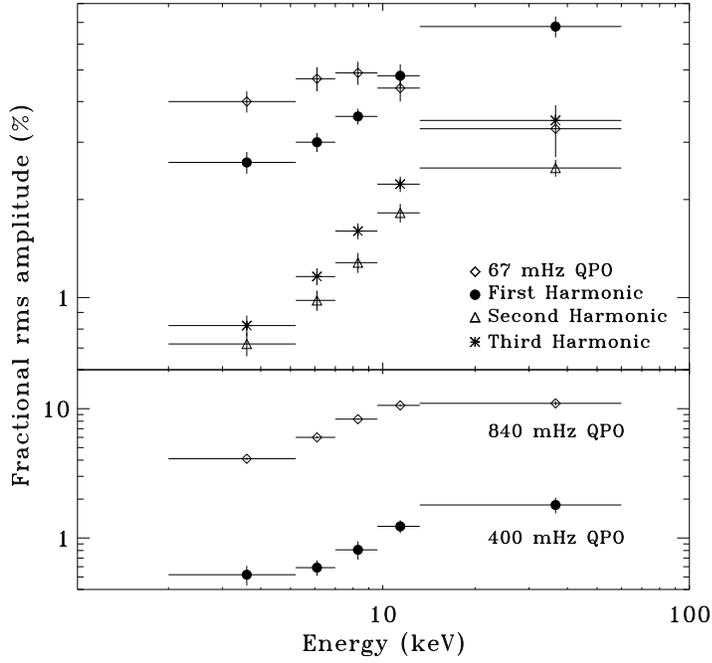,width=4in}
\caption{Energy dependence of QPO amplitude. Note that the 67 Hz QPO
is not shown here; the results can be found in Morgan et al. (1997). } 
\label{fg:rms}
\end{figure}

\begin{figure}
\psfig{figure=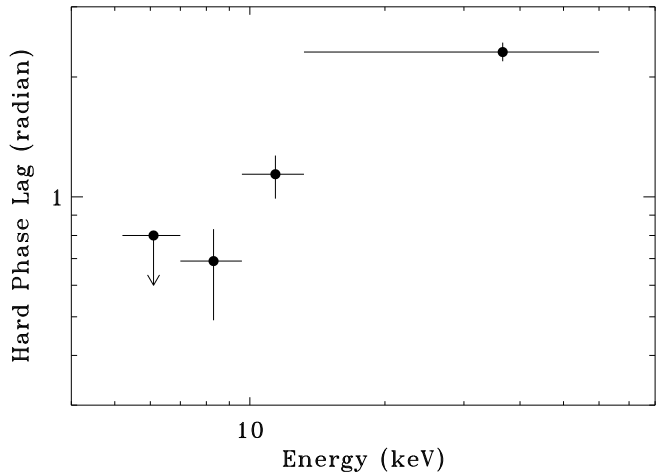,width=4in}
\caption{Energy dependence of hard phase lag associated with the 67
Hz QPO. The hard lag is not significantly detected in the lowest
energy band, so a $3\sigma$ upper limit is shown. }
\label{fg:lag67}
\end{figure}

\begin{figure}
\psfig{figure=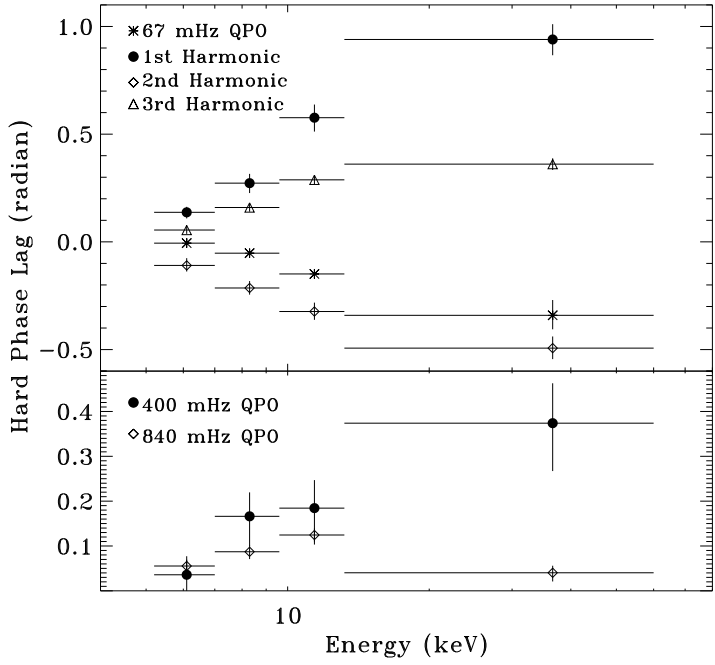,width=4in}
\caption{Similar to Fig.~3, but for phase lags associated with the 
low-frequency QPOs. Note that the negative values indicate soft lags. }
\label{fg:lag}
\end{figure}

\begin{figure}
\psfig{figure=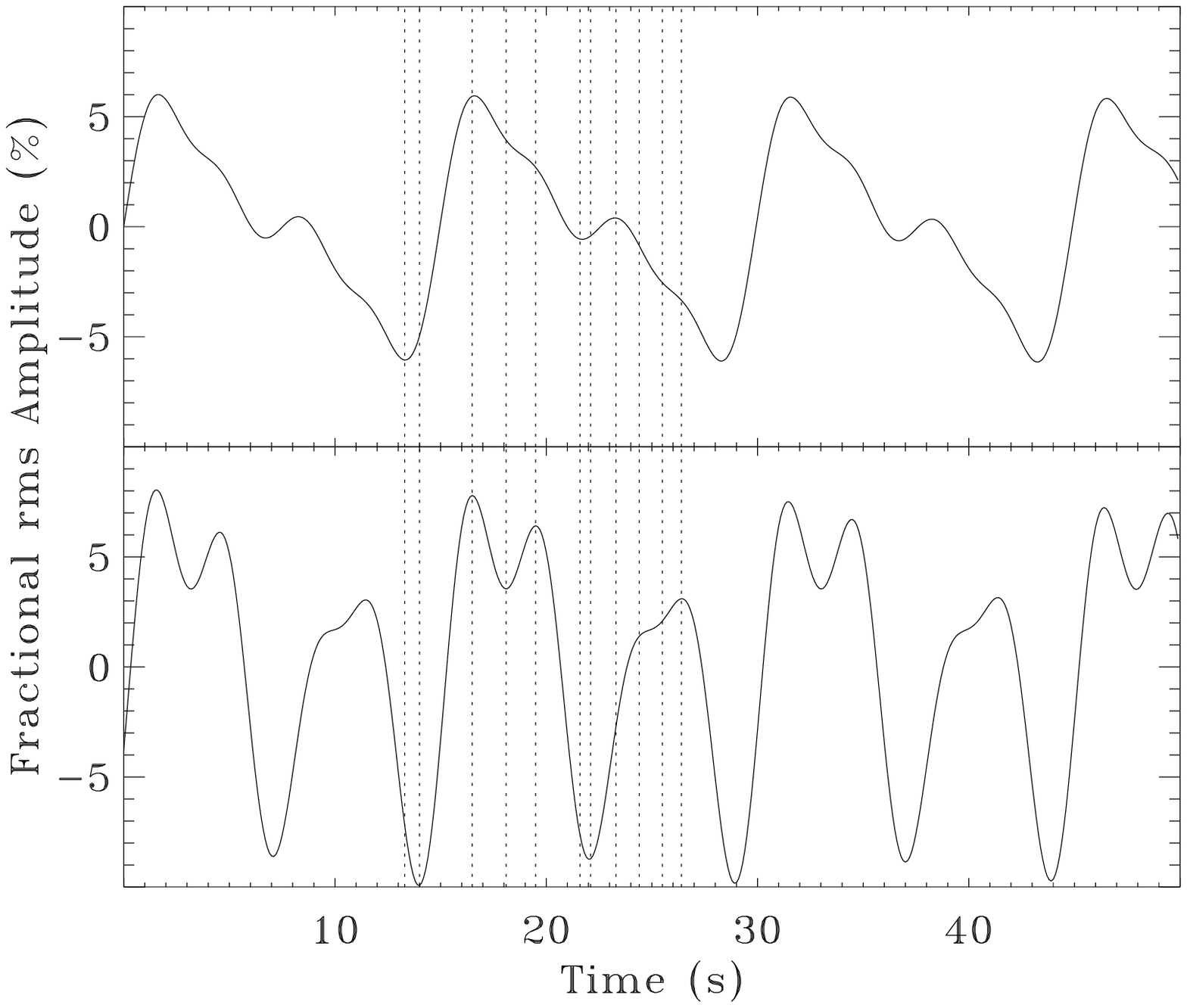,width=4in}
\caption{Simulated 67 mHz {\it periodic} signals with three harmonics
included. The simulation uses the actual fractional rms amplitude
and phase lag of each harmonic component of the 67 mHz QPO in two 
energy bands, 2--5.2 keV (top panel) and 13.2--60 keV (bottom panel). 
The dotted line marked all the transition points in each profile for
comparison. }
\label{fg:sim}
\end{figure}


\clearpage

\begin{references}
\reference{} B\"{o}ttcher,~M., \& Liang,~E.~P. 1998, \apj, 506, 281
\reference{} B\"{o}ttcher,~M., \& Liang,~E.~P. 1999, \apj, 511, L37
\reference{} Bradt,~H.~V., Rothschild,~R.~E., \& Swank,~J.~H. 1993,
\aaps, 97, 355 
\reference{} Castro-Tirado,~A.~J., Brandt,~S., \& Lund,~N. 1992,
\iaucirc\ 5590
\reference{} Chen,~X., et al. 1995, \apj, 443, L61
\reference{} Crary,~D.~J., et al. 1998, \apj, 493, 71
\reference{} Cui,~W., Zhang,~S.~N., Focke,~W., \& Swank,~J.~H. 1997,
\apj, 484, 383
\reference{} Cui,~W. 1997, talk at the Microquasar Workshop, Goddard 
Space Flight Center, Greenbelt, Maryland, May 1-3 
\reference{} Cui,~W., Zhang,~S.~N., \& Chen,~W. 1998, \apj, 492, L53
\reference{} Cui,~W. 1999, Proc. ``High-Energy Processes in Accreting
Black Holes'', eds. J. Poutanen \& R. Svensson (ASP: San Francisco),
ASP Conf. Ser. Vol. 161, P. 97 (astro-ph/9809408)
\reference{} Cui,~W., Zhang,~S.~N., Chen,~W., \& Swank,~J. 1999a, talk
at the 193rd American Astronomical Society Meeting, BAAS, vol. 30,
No. 4
\reference{} Cui,~W., Chen,~W., \& Zhang,~S.~N. 1999b, Proc. ``The 
Third William Fairbank Meeting on The Lense-Thirring Effect'',
eds. L.-Z. Fang \& R. Ruffini, in press (astro-ph/9811023)
\reference{} Dove,~J.~B., Wilms,~J., Maisack,~M., \&
Begelman,~M.~C. 1997, \apj, 487, 759
\reference{} Greiner,~J., Morgan,~E.~H., \&  Remillard,~R.~A. 1996,
\apj, 473, L107
\reference{} Hua,~X.-M., \& Titarchuk,~L. 1996, \apj, 496, 280
\reference{} Hua,~X.-M., Kazanas,~D., \& Cui,~W. 1999, \apj, 512, 793
\reference{} Kato,~S. 1989, \pasj, 41, 745
\reference{} Kazanas,~D., Hua,~X.-M., \& Titarchuk,~L. 1997, \apj,
480, 735
\reference{} Kazanas,~D., \& Hua,~X.-M. 1999, \apj, 519, 750
\reference{} Krolik,~J.~H., \apj, 498, L13
\reference{} Lampton,~M., Margon,~B., \& Bowyer,~S. 1976, \apj, 208,
177
\reference{} Lee,~H.~C., \& Miller,~G.~S. 1998, \mnras, 299, 479
\reference{} Lehr,~D.~E., Wagoner,~R.~V., \& Wilms,~J. 1999, BAAS, 31, 731
\reference{} Lightman,~A.~P., \& Eardley,~D.~M. 1974, \apj, 187, L1
\reference{} Luo,~C., \& Liang,~E.~P. 1998, \apj, 498, 307
\reference{} Manmoto,~T., Takeuchi,~M., Mineshige,~S., Matsumoto,~R.,
\& Negoro,~H. 1996, \apj, 464, L135
\reference{} Mirabel,~I.~F., \& Rodriguez,~L.~F. 1999, \araa, vol. 37,
in press (astro-ph/9902062)
\reference{} Miyamoto,~S., Kitamoto,~S., Mitsuda,~K., \&
Dotani,~T. 1988, Nature, 336, 450
\reference{} Morgan,~E.~M., Remillard,~R.~A., \& Greiner,~J. 1997,
\apj, 482, 993
\reference{} Narayan,~R., \& Yi,~I. 1994, \apj, 428, L13
\reference{} Nowak,~M.~A., Wagoner,~R.~V., Begelman,~M.~C., \&
Lehr,~D.~E. 1997, \apj, 477, L91
\reference{} Nowak,~M.~A., Vaughan,~B.~A., Wilms,~J., Dove,~J.~B., \&
Begelman,~M.~C. 1999, \apj, 510, 874
\reference{} Payne,~D.~G. 1980, \apj, 441, 770
\reference{} Poutanen,~J., \& Fabian,~A.~C. 1999, \mnras, 306, L31
\reference{} Tanaka,~Y., \& Lewin,~W.~H.~G. 1995, in ``X-ray
Binaries'', eds. W. H. G. Lewin, J. van Paradijs, \& E. P. J. van den 
Heuvel (Cambridge U. Press, Cambridge) p. 126
\reference{} van der Klis,~M. 1995, in ``X-ray Binaries'',
eds. W. H. G. Lewin, J. van Paradijs, \& E. P. J. van den Heuvel
(Cambridge U. Press, Cambridge) p. 252
\reference{} Wagoner,~R.~V., Silbergleit,~A.~S., Lehr,~D.~E., \& Ortega,~M.~A. 1999, BAAS, 31, 708
\end{references}
\end{document}